\documentclass{elsart}
\usepackage{epsfig,amssymb}

\begin{document}

\begin{frontmatter}
\title{\boldmath $\mu$SR in Ce$_{1-x}$La$_x$Al$_3$:\\ anisotropic Kondo effect?}
\author[UCR]{\underline{D. E. MacLaughlin},}
\author[UCR]{M.~S. Rose,}
\author[UCR]{B.-L. Young,}
\author[CSULA]{O. O. Bernal,}
\author[LANL]{R. H. Heffner,}
\author[KOL]{G. J. Nieuwenhuys,}
\author[UF]{R. Pietri}
and \author[UF]{B. Andraka}

\address[UCR]{Department of Physics, University of California, \\Riverside, California 92521-0413, USA}
\address[CSULA]{Department of Physics and Astronomy, California State University,\\ Los Angeles, California 90032, USA}
\address[LANL]{Los Alamos National Laboratory, Los Alamos, New Mexico 87545, USA}
\address[KOL]{Kamerlingh Onnes Laboratorium, Leiden University, \\2300 RA Leiden, The Netherlands}
\address[UF]{Department of Physics, University of Florida, Gainesville, Florida 32611, USA}

\begin{abstract}
Zero-field $\mu$SR experiments in the heavy-fermion alloys Ce$_{1-x}$La$_x$Al$_3$, $x = 0$ and 0.2, examine a recent proposal that the system exhibits a strong anisotropic Kondo effect. We resolve a damped oscillatory component for both La concentrations, indicative of disordered antiferromagnetism. For $x = 0.2$ the oscillation frequency decreases smoothly with increasing temperature, and vanishes at the specific heat anomaly temperature~$T^{\displaystyle*} \approx 2.2$~K\@. Our results are consistent with the view that $T^{\displaystyle*}$ is due to a magnetic transition rather than anisotropic Kondo behavior. 
\end{abstract}
\end{frontmatter}

The theory of the anisotropic Kondo model (AKM) has been studied extensively in recent years. In spite of this interest there have been few experimental studies of physical realizations of the AKM in systems of local d or f moments in metals, which were the earliest manifestations of the ``ordinary'' Kondo effect. 

The heavy-fermion compound~CeAl$_3$ was long thought to exhibit a nonmagnetic ``Kondo-lattice'' ground state. $\mu$SR~\cite{BOGH89} and NMR~\cite{GHO95} studies showed, however, that inhomogeneous weak-moment magnetic order sets in below $\sim$1~K\@. Moreover, as lanthanum is doped onto cerium sites the specific heat coefficient~$\gamma(T) = C(T)/T$ is drastically modified, and a maximum in $\gamma(T)$ at a characteristic temperature~$T^{\displaystyle*}$ ($\approx 0.4$~K in CeAl$_3$) moves up in temperature and grows into a large peak~\cite{AJS95}; $T^{\displaystyle*} = 2.2$~K for $x = 0.2$. This behavior was initially taken as evidence for development of a weak-moment magnetically ordered phase of CeAl$_3$, and attributed to reduction of the hybridization between Ce f electrons and ligand-derived conduction electrons. 

This interpretation has been called into question~\cite{GORM00} on the basis of inelastic neutron scattering experiments on Ce$_{0.8}$La$_{0.2}$Al$_3$, which found a broad inelastic peak below $T^{\displaystyle*}$. This peak was taken as evidence for applicability of the AKM to the Ce$_{1-x}$La$_x$Al$_3$ system. Zero-field $\mu^+$ spin relaxation (ZF-$\mu$SR) experiments~\cite{Sche85} at the ISIS pulsed muon facility~\cite{GORM00} indicated magnetic freezing at $T = T^{\displaystyle*}$, but the frozen moment was claimed to be weak (${\sim}0.05\ \mu_B$/Ce ion), principally because no neutron diffraction peak was observed. Recent ZF-$\mu$SR studies~\cite{GORC01}, also carried out at ISIS, conclude that for $x = 0.05$ the magnetic freezing persists to temperatures well above $T^{\displaystyle*}$. This is taken as further evidence that the specific heat peak at $T^{\displaystyle*}$ reflects the anisotropic Kondo effect and is unrelated to spin freezing. 

Specific heat measurements in Ce$_{0.8}$La$_{0.2}$Al$_3$ at applied fields of up to 14~T~\cite{PIA00} disagree with the theoretical behavior of the AKM, however. Furthermore, the argument for applicability of the AKM depends strongly on evidence that the frozen moments are small, viz., the absence of neutron Bragg peaks. Neutron diffraction would not necessarily be sensitive to spin freezing of a glassy nature or short-range order, however, whereas the ZF-$\mu$SR relaxation rate reflects spin freezing independently of the degree of long-range order.

\begin{figure}[t]
\begin{center}
\epsfig{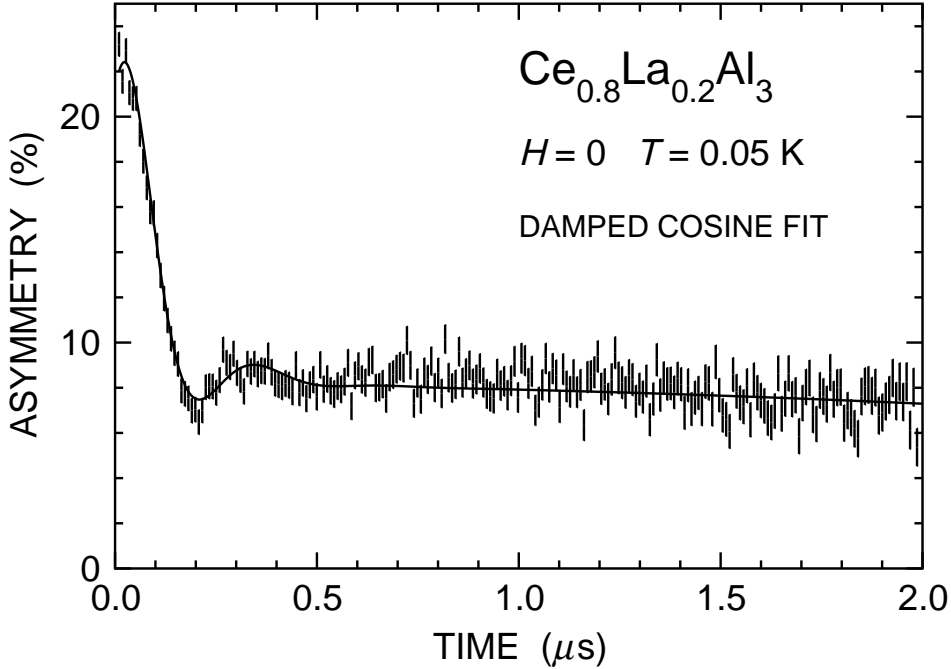}
\end{center}
\caption{Ce$_{0.8}$La$_{0.2}$Al$_3$ asymmetry relaxation function.}
\label{fig:CeLaAlfig1}
\end{figure}

We have carried out ZF-$\mu$SR experiments in Ce$_{1-x}$La$_x$Al$_3$, $x = 0$ and 0.2, at TRIUMF and PSI\@. For $x = 0.2$ we observe a strongly damped but resolved oscillation in the muon asymmetry data at low temperatures, as shown in Fig.~\ref{fig:CeLaAlfig1}. 
For both La concentrations the precession frequency disappears at $T^{\displaystyle*}$, in disagreement with the results of Ref.~\cite{GORC01} for $x = 0.05$, and in agreement with the expected result if $T^{\displaystyle*}$ were a magnetic ordering transition temperature. The strong damping indicates a distribution of static local fields at muon sites, as might be expected in a strongly disordered substitutional alloy, but the oscillation reflects a fairly well-defined average local field. For comparison with the analysis of Ref.~\cite{GORC01} asymmetry data for both La concentrations were fit assuming magnetic and nonmagnetic volume fractions; a ``damped cosine'' form was used for the precessing signal. The form of the relaxation function is
\begin{equation}
G(t) = A_m\left[ {\textstyle\frac{2}{3}}\exp(-\lambda_T t)\cos(2\pi\nu t + \phi) + {\textstyle\frac{1}{3}}\exp(-\lambda_L t) \right] + A_nG_{\rm KT}(t) \,.
\end{equation}
Here the relaxation from the magnetic volume fraction~$f_{\rm mag}$ is modeled by the component with asymmetry amplitude~$A_m$. The static field at the muon site is assumed to be randomly oriented. Relaxation of muons in the remaining nonmagnetic fraction is due to nuclear dipolar fields and is described by a static Kubo-Toyabe function~$G_{\rm KT}(t)$~\cite{Sche85}; thus $f_{\rm mag} = A_m/(A_m + A_n)$. The relaxation rates~$\lambda_T$ and $\lambda_L$ characterize the transverse and longitudinal relaxation, respectively; the former is expected to be dominated by static disorder whereas $\lambda_L$ reflects dynamic spin-lattice relaxation. The qualitative features of the fits are unchanged if the damped Bessel function found by Amato in CeAl$_3$~\cite{Amat97} is used.

\begin{figure}
\begin{center}
\epsfig{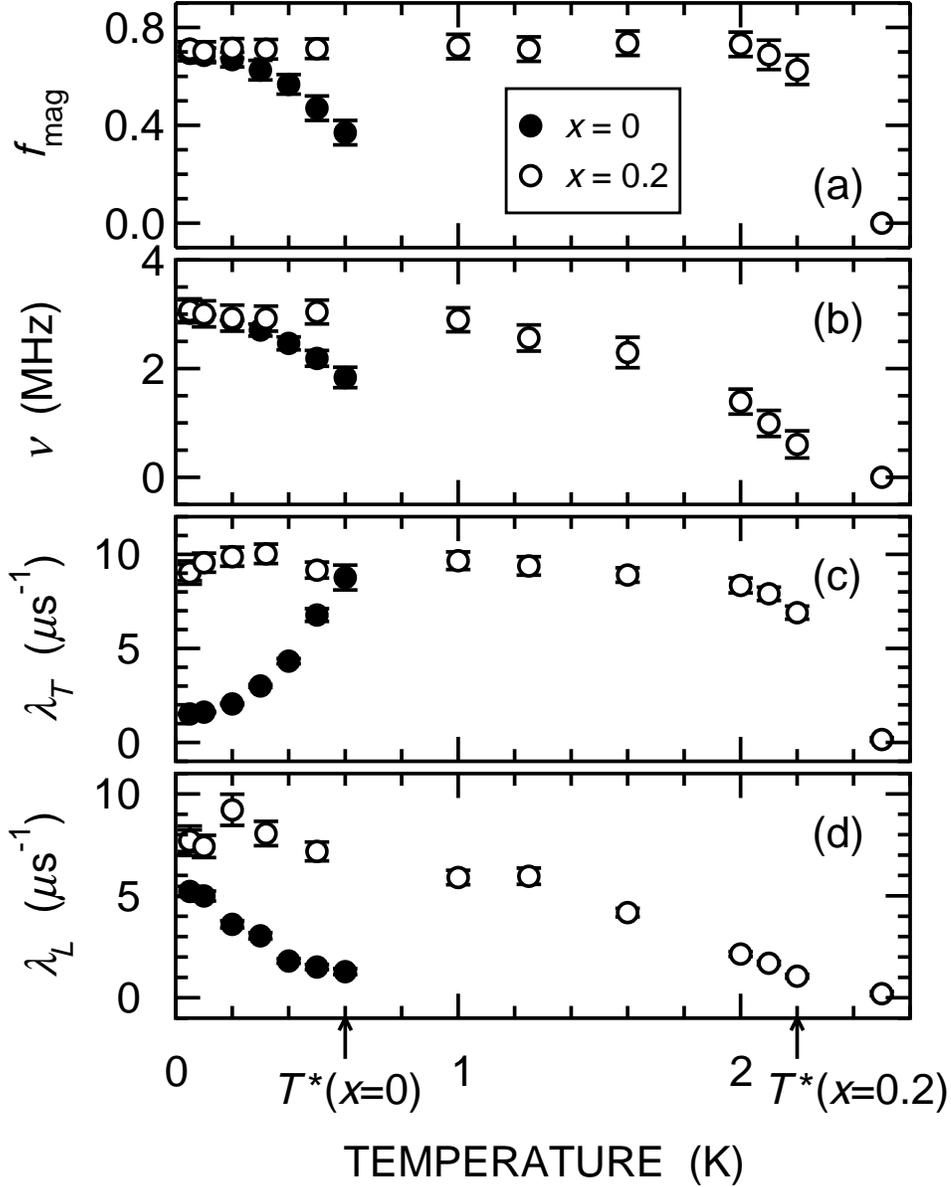}
\end{center}
\caption{Temperature dependence of (a) magnetic volume fraction~$f_{\rm mag}(T)$, (b) precession frequency~$\nu(T)$, (c) transverse relaxation rate~$\lambda_T(T)$, and (d) longitudinal relaxation rate~$\lambda_L(T)$ from ZF-$\mu$SR in Ce$_{1-x}$La$_{x}$Al$_3$, $x = 0$ and 0.2.}
\label{fig:CeLaAlfig2}
\end{figure}

The temperature dependence of the fit parameters is given in Fig.~\ref{fig:CeLaAlfig2}. 
It can be seen that $\nu(T)$ (a)~is independent of $x$ at low temperatures, (b)~decreases markedly as $T^{\displaystyle*}$ is approached from below, and (c)~is suppressed to zero at $T = T^{\displaystyle*}$ for $x = 0.02$. Unfortunately the muon stopping site in CeAl$_3$ is not known, otherwise the ordered moment~$\mu_{\rm ord}$ could be estimated accurately. Amato~\cite{Amat97} gives a range $\mu_{\rm ord} = 0.11$--0.5 $\mu_B$/Ce ion in CeAl$_3$, which is 2--10 times larger than the value reported in Ref.~\cite{GORM00}. The entropy release at $T^{\displaystyle*}$ implies $\mu_{\rm ord} \approx 0.3\ \mu_B$/Ce ion~\cite{AJS95}, comparable to values obtained from NMR and other measurements~\cite{NKAF88}. 

In CeAl$_3$ $\nu(T)$ is not completely suppressed to zero and $\lambda_T(T)$ increases rapidly at $T^{\displaystyle*}$. This is not necessarily a sign of a first-order transition since it can be understood as an effect of inhomogeneity in the transition temperature, as observed in La$_2$CuO$_4$~\cite{MVBdR94}. No such increase in broadening is observed in Ce$_{0.8}$La$_{0.2}$Al$_3$. CeAl$_3$ also exhibits a decrease of $f_{\rm mag}$ as $T^{\displaystyle*}$ is approached from below; this is also not seen in Ce$_{0.8}$La$_{0.2}$Al$_3$.

In conclusion, our ZF-$\mu$SR data are consistent with a scenario~\cite{AJS95,PIA00} in which the specific heat anomaly in Ce$_{1-x}$La$_x$A$_3$ is associated with the onset of static magnetism below $T^{\displaystyle*}$, with or without long-range order. In particular, there is no sign of such static magnetism above this temperature in Ce$_{0.8}$La$_{0.2}$Al$_3$.

We are grateful to A. Amato, C. Baines, M. Good, D. Herlach, B. Hitti, S. Kreitzman, P. Russo, and A. Savici for help with the experiments, and to E. Goremychkin, K. Ingersent, R. Osborn, and B. Rainford for useful discussions. This work was supported in part by the U.S. NSF, Grant nos. DMR-9731361 and DMR-0102293 (UC Riverside), DMR-9820631 (CSU Los Angeles), and DMR-0104240 (U. of Florida), and by the Netherlands NWO and FOM, and was performed in part under the auspices of the U.S. DOE.

\end{document}